\documentstyle[preprint]{jpsj}

\def\runtitle{Neutron Scattering Study of Phase Diagram of 
(Cu$_{1-x}$Mg$_x$)GeO$_3$}

\def\runauthor{Hironori {\sc NAKAO}}

\title {Neutron Scattering Study of Temperature-Concentration Phase Diagram 
        of (Cu$_{1-x}$Mg$_x$)GeO$_3$}
\author{Hironori {\sc Nakao}%
\footnote{Present address: Photon Factory, 
 Institute of Materials Structure Science,
 High Energy Accelerator Research Organization,
 Tsukuba, Ibaraki 305-0801. 
 E-mail: hironori.nakao@kek.jp }
, Masakazu {\sc Nishi}, Yasuhiko {\sc Fujii},
Takatsugu {\sc Masuda}$^1$,\\ Ichiro  {\sc Tsukada}$^{1,}$%
\footnote{Present address: Central Research Institute of 
Electric Power Industry, 2-11-1 Iwado Kita, Komae-shi, Tokyo 201-8511.}, 
Kunimitsu {\sc Uchinokura}$^1$,
Kazuma {\sc Hirota}$^2$, and Gen {\sc Shirane}$^3$}

\inst{Neutron Scattering Laboratory, Institute for Solid State Physics,
   University of Tokyo, 106-1 Shirakata, Tokai, Ibaraki 319-1106. \\
 $^1$Department of Applied Physics,
   University of Tokyo, 7-3-1 Hongo, Bunkyo-ku, Tokyo 113-8656.\\
 $^2$Department of Physics, Tohoku University, Sendai 980-8578.\\
 $^3$Department of Physics, Brookhaven National Laboratory, Upton,
   NY 11973-5000, USA.\\}
\recdate
{\today}

\abst
{
In doped CuGeO$_3$ systems, 
such as (Cu$_{1-x}$Zn$_{x}$)GeO$_3$ and Cu(Ge$_{1-x}$Si$_{x}$)O$_3$,
the spin-Peierls (SP) ordering ($T\le T_{SP}$)
coexists with the antiferromagnetic (AF) phase
($T\le T_N\le T_{SP}$). $T_{SP}$ decreases
while $T_N$ increases with increasing $x$
in low doping region. 
For higher $x$, however, the SP state disappears and 
only the AF state remains. 
These features are common for all the doped CuGeO$_3$ systems
so far studied, indicating the existence of universal $T-x$
phase diagram.
Recently, Masuda {\it et al.} carried out comprehensive
magnetic susceptibility ($\chi$) measurements of 
(Cu$_{1-x}$Mg$_x$)GeO$_3$,
in which doping concentration can be controlled significantly
better than the Zn doped systems.
They found that $T_N$ suddenly jumps
from 3.43 to 3.98~K at the critical concentration $x_c\sim 0.023$
and that a drop in $\chi$ corresponding to the SP ordering
also disappears at $x>x_c$.
They thus concluded that there is a compositional phase boundary
between two distinct magnetic phases.
To clarify the nature of two phases, we performed neutron-scattering
measurements on (Cu$_{1-x}$Mg$_x$)GeO$_3$ single crystals with various $x$.
Analysis of the data at fixed temperature points as a function of doping
concentration has revealed sudden changes of order parameters 
at the critical concentration $x_c=0.027\pm 0.001$. 
At finite temperatures below $T_N$, 
the drastic increase of the AF moment takes place at $x_c$.
The spin-Peierls order parameter $\delta$
associated with lattice dimerization
shows a precipitous decrease at all temperature below $T_{SP}$.
However, it goes to zero above $x_c$ only at the low temperature limit.
}
\kword
{ (Cu$_{1-x}$Mg$_x$)GeO$_3$, spin-Peierls transition, 
 antiferromagnetic transition, phase diagram, neutron scattering}

\begin{document}
\sloppy
\maketitle

\section{INTRODUCTION}
Since the  first inorganic spin-Peierls(SP) compound CuGeO$_3$ was
discovered,~\cite{hase} extensive studies have been performed to understand
the mechanism of the SP transition. 
In the course of studying doping effects on the SP state
using Cu(Ge$_{1-x}$Si$_x$)O$_3$,
Regnault {\it et al.}~\cite{regnault} found that the spin-Peierls (SP)
state ($T\le T_{SP}$) coexists with the antiferromagnetic (AF)
state ($T\le T_N\le T_{SP}$).
The discovery of coexistence has attracted much interest
because it is a novel spin system for studying competition
and cooperation between a singlet state with lattice
dimerization and a three-dimensional N\'eel state.
Fukuyama {\it et al.}~\cite{fukuyama1} have theoretically shown
that impurity-induced moments and lattice dimerization
are both true long-range orderings though they have spatial modulations
reflecting distribution of impurities.
Kojima {\it et al.}~\cite{kojima} observed that
antiferromagnetic ordering is indeed spatially modulated
by muon spin relaxation ($\mu$SR).

Two classes of doped systems are so far reported, i.e.
the site-random system such as (Cu$_{1-x}M_x$)GeO$_3$
($M=$~Zn,~\cite{manabe,sasago1,martin-zn} Mg~\cite{ajiro-mg,masuda-mg}) 
and the bond-random system such as 
Cu(Ge$_{1-x}$Si$_x$)O$_3$.~\cite{schoeff-si,hirota-si,katano-si}
Both systems have very similar temperature vs. doping-concentration
$T-x$ phase diagrams.
$T_{SP}$ decreases almost linearly from 14.2~K
at $x=0$ with increasing $x$.
The SP order parameter, i.e. lattice dimerization $\delta$,
also decreases as $x$ increases, 
and finally disappears at higher $x$.
On the other hand, the AF order takes place at $T_{\rm N}=28.5$~mK
even at $x=0.00112$, and gradually increases with increasing $x$.~\cite{manabe}
$T_N$ reaches the maximum temperature 
and disappears at the high concentration region.

Martin {\it et al.}~\cite{martin-zn} carried out comprehensive
neutron scattering experiments on Zn-doped CuGeO$_3$.
They reported that the SP phase exists in samples with
exceptionally  high concentrations of Zn.
They also pointed out that it is hard to dope Zn 
uniformly, particularly at high concentrations.
Masuda {\it et al.}~\cite{masuda-mg} reported 
that the doping concentration can be 
controlled significantly better in (Cu$_{1-x}$Mg$_x$)GeO$_3$,
and carried out comprehensive magnetic susceptibility ($\chi$)
measurements on high quality single crystals of (Cu$_{1-x}$Mg$_x$)GeO$_3$
to establish the $T-x$ phase diagram.
At $x=0.023$, they found two anomalies at $T=3.43$ and $3.98$~K,
which they ascribed to AF orderings, 
while another anomaly at $T=10$~K clearly indicates the SP transition,
as shown in Fig.~\ref{fig1}.
\begin{figure}[tbh]
 \vspace*{1cm}
 \caption{
 (Bottom) The temperature-concentration phase diagram of  
 (Cu$_{1-x}$Mg$_x$)GeO$_3$
 previously reported by magnetic susceptibility measurements [ref.~9]. 
 At $x<x_c$ the AF ordering is realized on a dimerized lattice (D-AF) 
 while the AF ordering is on a uniform lattice (U-AF) at $x>x_c$. 
 (Top) Raw data of susceptibility measured on the $x=0.023$.  
 Two $T_N$'s suggest coexistence of D-AF and U-AF phases near $x_c$.
 }
 \label{fig1}
\end{figure}
They reported that the $x$ dependence of $T_N$ clearly
exhibits a discontinuity at this concentration and 
a drop in $\chi$ corresponding to the SP ordering also
disappears at $x>0.023$.
From these results, they concluded 
that there exists a compositional phase boundary 
at critical concentration $x_c=0.023$
between two different magnetic phases.
At $x<x_c$, the AF ordering is realized on a dimerized lattice 
(D-AF) while the AF ordering on a uniform lattice (U-AF) at $x>x_c$.

To understand the nature of possible compositional transition and 
two different phases, we have performed neutron-scattering measurements
on (Cu$_{1-x}$Mg$_x$)GeO$_3$ single crystals with various $x$ values.
We indeed confirmed that the SP order parameter, i.e.
lattice dimerization $\delta$, exhibits a discontinuous change
at $x_c$ below $T_{SP}$,
which is clearly seen at low temperatures in particular.
We also found that the AF ordered moment at 1.3~K jumps.
For $x\ge x_c$, however, the SP order parameter $\delta$ remains
finite at finite temperatures though $\delta$ approaches zero at $T=0$~K.
The present results support the existence of compositional phase 
boundary at $x_c$ and the disappearance of SP ordering
for $x>x_c$ at $T=0$~K, though the SP phase at finite temperature
survives in some $T-x$ region above $x_c$.

\section{EXPERIMENTAL DETAILS}

A series of high-quality (Cu$_{1-x}$Mg$_x$)GeO$_3$ single crystals
($x=$~0.017, 0.026, 0.028, 0.032, 0.041, 0.082)  used in the present
experiment were grown by a floating-zone method.~\cite{masuda-mg}
A typical size of crystal was 20$\times$5$\times$3 mm$^3$, 
in which the longest direction was parallel to the orthorhombic $c$-axis. 
The concentration of Mg dopant, $x$, was carefully
determined by an inductively coupled plasma atomic emission
spectroscopy (ICP-AES), in which error is typically within
$\pm$0.001.\cite{masuda-mg}

Neutron-scattering experiments were carried out on ISSP-owned 
triple-axis spectrometers installed at the JRR-3M reactor of Japan Atomic
Energy Research Institute.  A crystal was mounted in an aluminum can with He
exchange gas, which was attached to a cold finger of 
an Orange cryostat capable of reaching 1.3~K. 
The sample was aligned so as to place the $(h\ k\ h)$ or $(0\ k\ l)$
zone in scattering plane. 
Incident neutron beams were monochromatized
with the $(0\ 0\ 2)$ reflection of pyrolytic graphite (PG) 
to select 14.7~meV or 5~meV
and the PG $(0\ 0\ 2)$ reflection was also used for analyzing neutron energy.
A double PG or Be filter was used to reduce higher-order
contaminations.  The beams were horizontally collimated,
with a typical configuration of 40'-40'-sample-40'-80' 
in sequence from the neutron source (reactor) to the detector.

The present $T-x$ phase diagram consists of traces of $T_{SP}(T,x)$ and
$T_{N}(T,x)$, which were determined at each $x$ using temperature
dependence of superlattice reflections
$I_{SP}(\frac{3}{2}\ 1\ \frac{3}{2})$ and $I_{AF}(0\ 1\ \frac{1}{2})$,
respectively.  Effective atomic displacement $\delta_{eff}$ 
for lattice dimerization in the SP state and effective magnetic moment
$\mu_{eff}$ in the AF state were evaluated by comparing several superlattice
intensities with fundamental reflections such as $(1\ 2\ 1)$ and $(0\ 2\ 1)$.
Typical lattice constants at room temperature are a=4.87~\AA, b=8.49~\AA, and
c=2.95~\AA\  for $x=0.032$.

\section{EXPERIMENTAL RESULTS}
\subsection{Phase transition}

Comprehensive neutron-scattering studies were carried out using a series of
(Cu$_{1-x}$Mg$_{x}$)GeO$_3$ single crystals in a wide concentration range
($x=0.017-0.082$).  For all the samples, we measured temperature dependence of
both the SP and AF superlattice reflections,
$I_{SP}(\frac{3}{2}\ 1\ \frac{3}{2})$ and $I_{AF}(0\ 1\ \frac{1}{2})$.
The magnetic susceptibilities were also measured for comparison.  Typical
results are shown for $x=0.017$ and $0.032$ in Figs.~\ref{fig2} and \ref{fig3},
respectively. 

\begin{figure}[tbh]
 \vspace*{1cm}
 \caption{
 Temperature dependence of the magnetic susceptibility (top) and 
 the neutron intensities of the superlattice peaks of 
 $I_{SP}(\frac{3}{2}\ 1\ \frac{3}{2})$ 
 and $I_{AF}(0\ 1\ \frac{1}{2})$ corresponding to the SP and AF phases, 
 respectively, (bottom) for the sample with $x=0.017<x_c$. 
 Both measurements clearly show $T_{SP}$ and $T_N$.
 }
 \label{fig2}
\end{figure}
\begin{figure}[tbh]
 \vspace*{1cm}
 \caption{
 Temperature dependence of the magnetic susceptibility (top) and
 the neutron intensities of superlattice peaks of 
 $I_{SP}(\frac{3}{2}\ 1\ \frac{3}{2})$
 and $I_{AF}(0\ 1\ \frac{1}{2})$ corresponding to the SP and AF phases, 
 respectively, (bottom) for the sample with $x=0.032>x_c$.  
 The susceptibility measurement shows no sign of  $T_{SP}$ 
 while neutron scattering data 
 clearly show a superlattice peak and a well-defined $T_{SP}$.
 }
 \label{fig3}
\end{figure}
For the sample with $x=0.017$, which is below the critical concentration
$x_{c}$, two sharp anomalies are seen in the susceptibility.
These temperatures are in good agreement with $T_{N}$ and $T_{SP}$ 
determined from the order parameters.  
In the $x=0.032$ sample, which is above $x_{c}$, the AF
transition is observed in both the susceptibility and
neutron-scattering measurements.  However, the SP transition is not noticeable
in the susceptibility data though the neutron scattering result clearly shows
the SP ordering.  Note that it is already difficult at
$x=0.023$ to define $T_{SP}$ by susceptibility measurement as shown in
Fig.~\ref{fig1}. 
Although Masuda {\it et al.}\cite{masuda-mg} concluded that
the SP phase completely vanishes at $x>x_{c}$ from their susceptibility
measurements, the present neutron-scattering study unambiguously demonstrates
that the SP phase survives in that region.  As we show later, however, the SP
phase at $x>x_{c}$ exhibits significant reduction in the order parameter,
broadening of the transition temperature, and decrease of the correlation
length, all of which indicate that there indeed exists a compositional phase
boundary at $x_{c}$ as Masuda {\it et al.}\cite{masuda-mg} pointed out.

In order to quantitatively evaluate transitions at $T_{SP}$ and $T_{N}$, 
neutron-scattering data were least-square-fitted with the following equation:

\begin{equation}
 I(T)=I_0 + \frac{I}{\sqrt{\pi}\Delta T}\int_T^{\infty}
\Bigl(\frac{\tau-T}{\tau}\Bigr)^{2\beta}
\exp\biggl[-\Bigl(\frac{\tau-T_{C}}{\Delta T}\Bigr)^2\biggr] d\tau ,
\label{eq:order}
\end{equation}
\begin{center}
($T_{C}$ = $T_{SP}$ or $T_{N}$)
\end{center}

\noindent
We have assumed a power-law description with a Gaussian distribution of
transition temperature as previously employed for a pure
CuGeO$_{3}$.~\cite{hirota-pure} Tables I and II summarize
the fitting parameters thus obtained 
for various concentrations, together with $T_{SP}$
and $T_{N}$ determined by susceptibility (in a column labelled $\chi$).
\begin{table}[tbh]
\caption{
Summary of physical quantities characterizing SP properties of
(Cu$_{1-x}$Mg$_x$)GeO$_3$ obtained from the present neutron scattering
data, except for the column designated by ($\chi$) which is $T_{SP}$ 
determined previously by magnetic susceptibility measurements[ref.~9]. 
See details in the text.
 }
\begin{tabular}{lclll} 
\hline
\multicolumn{1}{c}{$x$} & \multicolumn{2}{c}{$T_{SP}$ (K)} 
& $\Delta T$ (K)  & $\delta_{eff}$  \\ 
& $\chi$ & neutron & & \\
\hline
0.0    & - & 13.3 & & 1.00 \\
0.017  & 10.8 & 10.7 & 0.5 & 0.66(2) \\
0.026  & - & 8.8  & 0.9 & 0.57(5)\\
0.028  & - & 8.5  & 0.7 & 0.45(2) \\
0.032  & - & 8.0  & 1.1  & 0.28(1)\\
\hline
\end{tabular}
\end{table}
\begin{table}
\caption{
Summary of physical quantities characterizing AF properties of
(Cu$_{1-x}$Mg$_x$)GeO$_3$ obtained from the present neutron scattering 
data, except for the column designated by ($\chi$) which is $T_N$ 
determined previously by magnetic susceptibility measurements[ref.~9]. 
See details in the text.
}
  \begin{tabular}{lllll} 
\hline
\multicolumn{1}{c}{$x$} & \multicolumn{2}{c}{$T_N$ (K)} 
& $\Delta T$ (K) & $\mu_{eff}$ ($\mu_{\rm B}$) \\
& $\chi$ & neutron & & 
\\ \hline
0.017 & 2.7 & 2.4 & 0.3 & 0.17(1)  
\\
0.026 & 4.0, 3.4 & 3.2 & 0.4 & 0.28(1)  
\\
0.028 & 4.2 & 3.9 & 0.3 & 0.36(2)  
\\
0.032 & 4.2 & 4.2 & 0.2  & 0.34(1)   
\\
0.041 & 4.5 & 4.3 & 0.2  & 0.24(2)   
\\
0.082 & 2.1  & 1.9 & 0.2 & 0.16(1) \\
\hline
\end{tabular} 
\end{table}
The effective atomic displacement $\delta_{eff}$ at $T=T_{N}$ 
and magnetic moment $\mu_{eff}$ at $T\sim 0$~K 
are also listed.  $T_{N}$ values determined using Eq.~(\ref{eq:order}) are in
good agreement with those obtained from magnetic susceptibility data.

\subsection{Order parameter}

We now closely examine the two sets of order parameters, $I_{SP} \propto
\delta_{eff}^{2}$ and $I_{AF} \propto \mu_{eff}^{2}$, over a wide $x$ range in
order to construct the phase diagram from the neutron-scattering measurements. 
To scale the order parameters properly, 
a series of superlattice and fundamental reflections were collected 
for all the samples in the $(0\ k\ l)$ and $(h\ k\ h)$
zones for the AF and SP reflections, respectively. 
The results are summarized in Figs.~\ref{fig4} and \ref{fig5}.  
\begin{figure}[hbt]
  \vspace*{1.0cm}
 \caption{
 Temperature dependence of superlattice intensity $I_{SP}$,
 which is proportional to $\delta_{eff}^2$, for various concentrations.
 The values are normalized by $I_{SP}$ at $T=0$~K for a pure system $(x=0)$. 
 Solid curves are guides for the eye. 
 For $x>x_c$, $\delta_{eff}^2 (T=0)$ becomes nearly
 zero while for $x<x_c$  it remains finite.
 }
 \label{fig4}
\end{figure}
\begin{figure}[hbt]
 \vspace*{1.0cm}
 \caption{
  Temperature dependence of the magnetic superlattice intensity $I_{AF}$,
  proportional to $\mu_{eff}^2$, for various concentrations.
  The values are normalized by the (0\ 2\ 1) fundamental reflection. 
  Solid curves are guides for the eye. 
 }
 \label{fig5}
 \vspace*{-1.0cm}
\end{figure}
Note that the SP
intensity significantly decreases as $x$ increases.

A very simple yet clear demonstration of the phase boundary at $x_{c}$ is
a plot of the intensity data as a function of $x$ at a given temperature.  As
shown in Fig.~\ref{fig6}, the $x$ dependence of $\mu_{eff}^{2}$ at 1.5~K and
3.0~K as well as $\delta_{eff}^{2}$ at 1.3~K and 7.0~K clearly indicate that
there exists a critical concentration between $x=0.026$ and $0.028$, 
though we have rather small number of $x$ points; only five. 
\begin{figure}[tbh]
 \vspace*{1cm}
 \caption{
  Concentration dependence of (a) $T_N$, (b)$\mu_{eff}^2$ 
  at $T=1.5, 3.0$~K, and (c) $\delta_{eff}^2$ at $T=1.3, 7.0$~K. 
  All data clearly show abrupt changes at the critical
  concentration $x_c=0.027\pm 0.001$, 
  evidence for the compositional phase boundary at $x_c$.
 }
 \label{fig6}
\end{figure}
The critical concentration
can be also seen in the $x$ dependence of $T_{N}$.
These plots lead to a simple picture that the AF order parameter is 
suddenly enhanced at $x_{c}$, accompanied by a precipitous drop of 
the lattice dimerization.  
However, $\delta$ remains finite at finite temperature and
approaches zero only in the low temperature limit.  (See Fig.~\ref{fig4}.)

The present neutron-scattering measurements have confirmed the existence of a
compositional phase boundary as first reported by Masuda {\it et
al.}\cite{masuda-mg}  However the critical concentration $x_c$ is
determined to be $0.027\pm 0.001$, 
which is somewhat larger than $x_{c}=0.023$ previously
reported from susceptibility measurements.~\cite{masuda-mg}  This difference is
most likely caused by coexistence of two phases in the vicinity of $x_{c}$. 
The order parameter $I_{AF}$ measured by neutron scattering is dominated by the
majority phase, while susceptibility data may show two anomalies
corresponding to the majority and minority phases 
if the latter has enough volume.  
The susceptibility data for $x=0.023$ in Fig.~\ref{fig1} 
shows two AF anomalies. 
The anomaly at higher temperature, however, exhibits a tiny change while the
other shows a large drop, 
indicating that the lower $T_{N}$ is still dominant at $x=0.023$. 
Therefore we believe that the actual compositional phase boundary
is located higher than $x=0.023$, 
namely $x_{c}=0.027\pm 0.001$ as determined from
the present neutron-scattering study.  

\subsection{Line-broadening}

Throughout the whole range of concentrations studied, 
the AF phase shows long-range ordering; 
the superlattice peaks are almost resolution-limited.  
On the other hand, the SP phase shows a slight line-broadening 
at high concentrations beyond $x_{c}$. 
Figure~\ref{fig7}(a) shows a typical peak profile of
SP reflection measured around $(\frac{3}{2}\ 1\ \frac{3}{2})$ for $x=0.032$ at
$T=5$~K.  The instrumental resolution was experimentally determined using the
$(3\ 2\ 3)$ fundamental peak employing $\frac{\lambda}{2}$ incident neutrons. 
It is clear that the SP reflection shows a significant broadening 
along the $[0\ k\ 0]$ direction. 
\begin{figure}[tb]
 \vspace*{1.0cm}
 \caption{
 (a) The peak profile of $I_{SP}(\frac{3}{2},1,\frac{3}{2})$ 
 plotted with closed circles measured in the direction of $[0\ k\ 0]$ 
 at $T=5$~K for $x=0.032>x_c$ is significantly broader than 
 the experimental resolution represented with open circles.  
 (b) The inverse correlation length plotted against temperature.
 }
 \label{fig7}
\end{figure}
Note that such a broadening is not obvious in the $[h\ 0\ h]$ direction, 
indicating anisotropic correlation lengths; the correlation 
length $\xi_{[h0h]}$ is at least 1.5 times longer than  $\xi_{[0k0]}$.

Such a remarkable broadening along $[0\ k\ 0]$ is observable 
in a wide temperature range as shown in Fig.~\ref{fig7}(b), 
where the intrinsic line-width, i.e. the inverse correlation length $\kappa$
is plotted  as a function of temperature.   Note that the development of a
long-range SP order is considerably suppressed below $T_{N}$, which is
consistent with the decrease of $I_{SP}$ in the same temperature range. 
At present the data on line widths are limited.
They seem to suggest that the correlation length $\xi$ doesn't have an anomaly 
at the critical concentration $x_{c}$, but changes continuously.\cite{nakao}
A similar line broadening was already reported by 
Martin {\it et al.}~\cite{martin-zn} in Zn-doped CuGeO$_{3}$ system.
However, there were problems concerning inhomogeneity in Zn-doped samples,
which makes quantitative discussions difficult.

\section{DISCUSSION}
As shown in the $T-x$ phase diagram of (Cu$_{1-x}$Mg$_x$)GeO$_3$
constructed by present neutron-scattering study (Fig.~\ref{fig8}), 
one can clearly see the existence of a compositional phase boundary 
at the critical concentration $x_{c}$ .
\begin{figure}[t]
 \vspace*{1.0cm}
 \caption{
  The $T-x$ phase diagram of (Cu$_{1-x}$Mg$_x$)GeO$_3$ 
  based on the present neutron data. 
  The shaded region represents a short-range ordered SP state newly observed. 
  The critical concentration is nearly independent of temperature
  as shown with an obliquely-dashed line.
  }
 \label{fig8}
\end{figure}
Saito\cite{saito2} has theoretically studied impurity effects 
in a disordered spin-Peierls system and suggested 
that the long-range SP order disappears 
above a critical concentration of impurity. 
The lattice dimerization $\delta$ extrapolated to 
$T=0$~K (See Fig.~\ref{fig6}(c)) as a function of concentration $x$ is 
consistent with the theoretical prediction. 
At present, the theory discusses only the ground state properties. 
We believe that the ``intermediate'' phase indicated by a shaded region,
where the short-range ordered SP state exists,
will be also clarified by extending the theory to finite temperatures.

Figure~\ref{fig9} shows a relation between $T_{N}$ 
and the effective magnetic moment $\mu_{eff}$ of 
(Cu$_{1-x}$Mg$_{x}$)GeO$_{3}$ at the low temperature limit. 
\begin{figure}[tbh]
 \vspace*{1.0cm}
 \caption{
  The effective magnetic moment $\mu_{eff}$ of (Cu$_{1-x}$Mg$_x$)GeO$_3$ 
  evaluated at the low temperature limit ($T=1.3$~K) plotted 
  against $T_N$ (closed circles).  
  Previously reported data on (Cu$_{1-x}$Zn$_x$)GeO$_3$ [ref.~7] 
  and Cu(Ge$_{1-x}$Si$_x$)O$_3$ [ref.~11, 12] are also shown.
  }
 \label{fig9}
\end{figure}
Data for Zn-doped and Si-doped CuGeO$_{3}$ are also shown. 
We notice that $\mu_{eff}$ increases linearly as $T_{N}$ increases, 
though the gradient seems to depend upon dopant. 
Note that the anomaly of $\mu_{eff}$ at $x_{c}$ is 
not very clear in this plot. 
This is because the anomaly in $\mu_{eff}$ is accompanied
by the jump of $T_{N}$ at $x_{c}$.

It is now clear that there exist two types of AF state 
separated by the compositional phase boundary at $x_{c}$. 
However, the origin of such a difference is still not completely understood. 
It is interesting to study how the difference between the dimerized (D-AF)
and undimerized AF (U-AF) states appears in the magnetic excitations. 
In addition to performing a comprehensive study of spin waves 
at different $x$, we plan to investigate how the structural dynamics, 
i.e. phonon and phason, affect the magnetic excitations 
by employing polarized neutrons.

In conclusion, the present neutron-scattering study of 
(Cu$_{1-x}$Mg$_{x}$)GeO$_{3}$ with a wide range of $x$ has revealed 
that the order parameters such as the lattice dimerization $\delta$ 
and the AF magnetic moment show a sudden change at the critical 
concentration $x_{c}=0.027 \pm 0.001$.  
This indicates the existence of a compositional phase boundary 
between two distinct AF phases. 
We also found that there exists an intermediate phase 
with a short-range SP order at $x>x_{c}$ in a certain finite temperature range.

\acknowledgement
We acknowledge M. Saito and H. Fukuyama for fruitful discussions.
We thank S. Watanabe for technical support on our experiments.
Two of authors (H.N. and T.M.) acknowledge support by the Japan
Society for the Promotion of Science for Young Scientists.
This study was supported in part by the U.S.-Japan Cooperative
Program on Neutron Scattering between USDOE and MONBUSHO
and by NEDO (New Energy and Industrial Technology Development Organization)
International Joint Research Grant.

\end{document}